\documentclass[10pt,a4paper]{article}
\usepackage{slashed}

\usepackage{amssymb}
\newtheorem{thm}{Theorem}

\newtheorem{lem}{Lemma}
\usepackage{amssymb}
\usepackage{amsmath}
\usepackage{color}
\usepackage{colordvi}

\title{ The phase of the scattering operator\\
from the geometry of certain infinite-dimensional groups}
\author{Jouko Mickelsson\\Department of Mathematics and Statistics, University of Helsinki}
\begin{document}

\maketitle

\begin{abstract} We revisit the computation of the phase of the Dirac fermion scattering operator
in external gauge fields. The computation is through a parallel transport along the path
of time evolution operators. The novelty of the present paper compared with the earlier
geometric approach by Langmann and Mickelsson \cite{LM} is that we can avoid the somewhat
arbitrary choice in the regularization of the time evolution for intermediate times using
a natural choice of the connection form on the space of appropriate unitary operators.
\end{abstract}

\section{Introduction} 

In quantum mechanics the scattering operator $S$ for Dirac fermions in an external vector
potential $A$ is computed from the asymptotics of the unitary time evolution operator $U(t),$
$$i\frac{\partial}{\partial t} U(t) = D_A U(t)$$
where $D_A$ is the Hamilton operator and the initial condition is $U(0)=1,$ assuming that the potential $A$ is switched off for times $t<0.$ The scattering operator is then $S = \text{lim}_{t\to\infty}  U_0(t)^{-1} U(t)$ and this limit exists
(in strong operator topology) when the potential is smooth and goes to zero enough rapidly
at infinity. Here $U_0(t)$ is the free time evolution corresponding to $A=0$ with the
initial condition $U_0(0)=1.$ 

In quantum field theory one needs to promote $S$ to an unitary operator $\hat S$ in the fermionic Fock space. According to the Shale-Stinespring \cite{ShSt} theorem this is possible when the off-diagonal blocks of
$S$ in the energy polarization (with respect to the free Dirac hamiltonian $D_0$) are Hilbert-Schmidt,
and this is indeed the case under the above mentioned restrictions on the potential $A.$ The unitary
operators with this restriction for a \it restricted unitary group, \rm denoted by $U_{res}.$
This groups and its representation theory was studied in detail in \cite{PrSe}.
The only problem is that the phase of $\hat S$ is not determined by the canonical quantization
procedure. 

The physics question is: Why bother about the phase? Physicists know how to handle these things
in perturbation theory since the work of Richard Feynman, Julian Schwinger and Sin-Itoro
Tomonaga (in the
case of quantum electrodynamics) around 1950. The renormalizability of the perturbation series,
in terms of Feynman integrals, has later been extended to weak interactions and to QCD. 
The point here is that the external field problem is essentially the only situation for 
realistic particle physics models where in principle the solution should be written down
in a nonpertubative way. By external field problem I mean here the quantization of the Dirac
field but keeping the gauge field classical; in perturbation series only the Feynman propagator
for the fermion field appears. 

Even in the external field problem there are diverging (1-loop) Feynman diagrams, and these contribute to
the phase of the scattering operator. The phase is not purely an academic question since it 
leads to a modification of the effective action and this in turn modifies the gauge field
propagator, which in the case of QED leads to the experimentally very precisely
measured modification of certain hydrogen atom energy levels (Lamb shif, \cite{LR}). However,
the 1-particle scattering operator $S$ satisfies the technical condition (see below) needed
so that the operator can be promoted to an unitary operator $\hat S $ in the fermionic
Fock space.

In physics the quantum scattering operator is computed using the Feynman rules, the matrix elements
are (nonconvergent) sums in the perturbation series, the individual terms given by Feynman 
diagrams. What is worse, some of the Feynman integrals are diverging, one has to introduce 
some renormalization methods to subtract the divergent parts.
But for the above problem this is not very satisfactory situation since in principle
the scattering operator $\hat S$ should be well-defined, convergent.

Denoting $U^I(t) = U_0(-t) U(t)$ one  could try to determine the phase by a parallel tranport,
provided that $U^I(t)$ is for all times $t$ in a suitable infinite-dimensional group. One
candidate for such a group is $U_{res}(H_+\oplus H_-)$ for a polarized Hilbert space $H.$ 
Here $H$ is the Hilbert space of square-integrable fermion fields and the polarization is defined
by the sign of the free hamiltonian $D_0.$ This is natural idea since the operator $S$ is actually
in $U_{res}.$ However, for finite times $t$ the time evolution $U^I(t)$ is not in $U_{res}.$ 

The phase problem has earlier been discussed from different points of view, \cite{SchFi}, \cite{Sch},
\cite{GbV},
\cite{DDMS}, \cite{LM}, \cite{Mi98}. A comparison of the different approaches is given in \cite{Laz}.
The aim of the present paper is to remove a defect in the earlier work \cite{LM} which is due to
an arbitrary choice of a regularization of the 1-particle time evolution for the intermediate
times, in order to bring the time evolution operators to $U_{res}$ in the geometric approach
proposed in [LM]. The different choices lead to different phases of the scattering operator,
the change being given by a holonomy along along a closed loop in $U_{res}.$ 

In this paper I want to show that there is a natural way to define the parallel transport on
an appropriate space of unitary operators (which includes $U_{res}$) which makes the result
independent of the choice of the regularization. The essential ingredient is the splitting of the
operators in the parallel transport formula to trace class operators and operators which are
conjugate to pseudodifferential operators. Then one can apply a generalized trace calculus
to these operators which, when applied to the Feynman diagrams, would be equivalent to the
dimensional regularization; in particular, the process eliminates the logarithmic divergencies
by subtracting an infinite quantity related to the Guillemin - Wodzicki residue of the operator.

The plan of the paper is the following. In Section 2 the basic geometrical properties of
the restricted unitary group $U_{res}$ are recalled. The bulk of the paper is the Section 3
containing the main result (Theorem 1) which shows that there exist a connection defining
the parallel transport on a space of unitary operators containing the time evolution operators;
this connection defines the phase of the quantum scattering operator. In section 4 the geometric
phase is compared with the 1-loop perturbation theory and seen to agree with the dimensional
regularisation.

Writing this paper was inspired by many discussions with Dirk Deckert, Detlef D\"urr, Franz Merkl, and 
Martin Schottenloher, and later with Jos{\'e} Gracia-Bond{\'i}a and Joseph V{\'a}rilly; I want to thank for the
invitation to visit LMU in M\"unich in September 2009 and University of Zaragoza in October
2010. The completion of the paper was unfortunately 
delayed for a long time because of other duties and interests.

\section{Parallel transport on the group $U_{res}$}

Let $\hat G$ be a central extension of a Lie group $G$ by
$\Bbb C^{\times}.$ The Lie algebra $\hat{\frak{g}}$ of $\hat G$ is a vector
space direct sum $\frak{g}\oplus\Bbb C.$ Let $\pi$ be the projection on
the second summand and let $\theta=dgg^{-1}$ be the right invariant
Maurer-Cartan one-form. We can then define a complex valued one-form $\phi$ on
$\hat G$ by $\phi=\pi(\theta).$ This is a connection form in the
principal $\Bbb C^{\times}$ bundle $\hat G\to G.$ Its curvature is a
left invariant two-form on $G$ given by $\omega(X,Y)=c(X,Y),$ where
left invariant vector fields $X,Y$ on $G$ are identified as elements
of the Lie algebra and $c$ is the 2-cocycle on $\frak{g}$ defining the
central extension,
$$ [(X,\lambda),(Y,\mu)]=([X,Y],c(X,Y)).$$

Let   $GL_{res}$  be the group of invertible linear transformations
$g:H\to H$ such that $[\epsilon,g]$ is Hilbert-Schmidt so that  $U_{res}$ is its
unitary
subgroup.  Let us apply the above remarks to $G=GL_{res},$ and to the Lie
algebra cocycle $c$  arising when promoting the one-particle
operators to operators  in the fermionic Fock space.

The central extension $\widehat{GL}_{res}$ is a nontrivial $\Bbb C^{\times}$
bundle over the base $GL_{res},$ \cite{PrSe}.
The elements of the group $\widehat{GL}_{res}$ (containing the unitary
subgroup $\hat  U_{res}$) can be thought of equivalence classes of pairs
$(g,q),$  where $g\in GL_{res}$ and $q:H_+\to H_+$ is an invertible
operator such that $a-q$ is a trace-class operator,
$$g=\left(\begin{matrix} a&b\\c&d\end{matrix}\right).$$
We have assumed that ind$\,a=0.$ If this is not the case, the subspace
$H_+$ must be either enlarged or made smaller by a suitable
finite-dimensional subspace in order to achieve ind$\,a=0.$
The equivalence relation is determined by $(g,q)\sim (g',q')$ if
$g=g'$ and det$(q'q^{-1})=1.$ Thus the fiber of the extension is $\Bbb C^
{\times}$ and it is parameterized by (the nonexisting ) determinant of $q.$
Here we can restrict to the Fredholm index ind$\,a =0$ subgroup since 
the continuous time evolution starting from the identity operator implies
that $U(t)$ for all $t$ is in the connected component of the identity.

The product is defined simply $(g,q)(g',q')=(gg',qq').$ Near the unit
element in $GL_{res}$ we can define a local section $g\mapsto (g,a),$
[PrSe].
Denoting
$$g^{-1}=\left(\begin{matrix} \alpha & \beta \\ \gamma & \delta \end{matrix}\right)$$
we can write the connection form as
\begin{equation} \phi_{g}\,=\text{tr}\, [(dgg^{-1})_a- dqq^{-1}]
=\text{tr}\, [ da\alpha + db\gamma -dq q^{-1}].  \end{equation}
The curvature of this connection at $g=1$ is
$$\omega= -\text{tr}\, (db dc ).$$
Interpreting the tangent vectors at $g=1$ as elements in the Lie algebra
we obtain
$$\omega(X,Y) = -\text{tr}(b(X) c(Y) -b(Y) c(X))= \frac14\text{tr}\, \epsilon[\epsilon,
X][\epsilon,Y].$$
Thus the curvature of the connection is directly given through the
Lie algebra central extension as promised.

We compute the parallel transport determined by the connection in the
range of the local section. Let $g(t)$ be a path in $GL_{res},$
$0 \leq t\leq T,$ with $g(0)=1.$ The lift $(g(t),q(t))$ is parallel if
\begin{equation}  0= \phi_{g(t),q(t)}(dg,dq)= \text{tr} [a'(t)\alpha(t)
 +b'(t) \gamma(t) - q'(t)q(t)^{-1}]. \end{equation}
Thus the parallel transport, relative to the trivialization $g\mapsto
(g,a),$ along the path $g(t)$ in the base is
accompanied with the multiplication by the complex number
\begin{equation} \exp\{-\int_{0}^T\text{tr} [a'(t) (\alpha(t)-a(t)^{-1})+b'(t)\gamma(t)]dt
\} \end{equation}
in the fiber $\Bbb C.$

Formally,
$$\text{tr}\, q'q^{-1}=\text{tr}[a'\alpha+b'\gamma]$$
and so
$$ \text{det} \, q(T) = \exp\int_{0}^T \text{tr}[a'(t)\alpha(t)+b'(t)\gamma(t)]dt$$
and also
$$ \text{det}\,a(T)= \exp\int_{0}^T \text{tr} \,a'(t)a(t)^{-1} dt.$$
Individually, the traces in these two expressions do not converge,
but putted together the trace converges and gives
\begin{equation} \text{det}(a(T)q(T)^{-1})=\exp\{ \int_{0}^T
\text{tr} [(a'(\alpha -a^{-1})+b'\gamma]dt\}. \end{equation}
Note that the exponent diverges outside of the domain of the local section,
reflecting the fact that det$\,a(T)=0$ outside of the domain.

\section{Time evolution of fermions in external gauge fields}

We shall study massless Dirac fermions  coupled to a gauge potential $A$
in Minkowski space. The potential is a smooth 1-form $A_{\mu}(x)\, dx^{\mu}$ in
space-time with
values in the Lie algebra $\mathfrak g$ of a compact gauge group $G.$ The elements
of $\mathfrak g$ are represented by hermitean (according to physics literature
convention) matrices in the complex vector space $\Bbb C^N.$ The free Dirac
operator is then $i\sum_{\mu=0}^{d}\gamma^{\mu} \partial_{\mu} +m.$ The metric is $x^2 = g_{\mu\nu} x^{\mu} x^{\mu}=
x_0^2 -x_1^2 - \dots -x_d^2.$ The Dirac gamma matrices satisfy
$\gamma_{\mu}\gamma_{\nu}
+\gamma_{\nu}\gamma_{\mu} = 2 g_{\mu\nu},$ $\gamma_0$ is hermitean and $\gamma_k$ is
antihermitean for $k\neq 0.$ 
The Dirac hamiltonian in the background gauge field $A$ is
$$D_A = -\gamma^0\sum_{k=1}^{d} \gamma^k(i\partial_k +A_k) -A_0 - \gamma_0 m,$$
where $m\geq 0$ is the mass of the fermion.
We shall take as the initial condition, at $t=0,$
 $A(\bold x,t=0)=0.$
We shall assume that $A(x)$ and its derivatives vanish faster than
$|x|^{-d/2}$ when $|x| \to \infty.$

The unitary time evolution operator of Dirac fermions in an external gauge potential $A$  is given as the solution of
$$ i\partial_t U(t) = (D_0 +V) U(t)$$
with the initial condition $U(0)=1$
where $D_0= -i\gamma^0 \gamma^k \partial_k -\gamma_0 m$  and $V= -i \gamma^0 \gamma^k A_k - A_0$  is the 
interaction term.  Even in the case of QED the time evolution $U(t)$ is not in the group
$U_{res},$ except in the case of pure electric field \cite{Rui}, \cite{Pal}.
According to \cite{Mi94}, \cite{LM}, we can choose an unitary operator 
$T_A$ which depends on $A$ and its time and space derivatives such that the renormalized
time evolution $U_{ren}(t) = T_A U(t)$ satisfies the differential equation
$$i\partial_t U_{ren}(t) = (D_0 +V_{ren}) U_{ren}(t),$$
where 
$$ V_{ren} = T_A^{-1} V T_A + T_A^{-1} i\partial_t T_A$$ 
is such that the commutator $[\epsilon, V_{ren}]  \in L_2.$ Here $\epsilon= \epsilon(D_0)$ is defined through the spectral step function
$\epsilon(x)= -1$ for $x<0$ and $\epsilon(x) = +1$ for $x\geq 0.$
Actually, it is possible to 
choose $T_A$ such that the commutator is even a trace-class operator [LM].  The operator
$T_A$ is a pseudodifferential operator which differs from the unit operator by a pseudodifferential
operator of order $-1.$

Going to the interaction picture,
$ U^I_{ren} = e^{itD_0} U_{ren},$ we have 
$$ i\partial_t U^I_{ren} (t) = V_{ren}^I U^I_{ren}(t)$$
with $V^I_{ren} = e^{itD_0} V_{ren} e^{-itD_0}.$ Now the interaction $V^I_{ren}$ is an element of
the Lie algebra $\mathfrak{u}_1$ of the group $U_1(H)$ consisting of unitaries in $H=H_+\oplus H_-$ with trace-class off-diagonal blocks
and is a continuous function of
the time $t.$ It follows that the time evolution equation has a differentiable (in time) solution
$U^I_{ren}(t).$

We can now apply the above results to the 'renormalized' one-particle
time evolution operators $g(t)= U^I_{ren}(t)$ in the interaction picture.
Let us, for the sake of simplicity, assume that the interaction is
switched off outside of a finite interval $[0,T]$ in time. Thus
the 1-particle scattering operator is $S_A = U^I_{ren}(T) = U(T).$
For all times $t$,
$g(t)\in  U_1.$ On the other hand, in the Fock representation of
$\widehat{GL}_1$ these correspond to elements $\hat g(t)$ in the
central
extension $\hat  U_1.$ The phase of the quantum time evolution operator
is then uniquely given by the parallel transport described above.

The Minkowskian effective action $Z(A)$ is by definition the vacuum expectation
value of the quantum scattering operator  $\hat S_A.$ The vacuum is invariant
under the free time evolution $\exp(itD_0)$ and taking into account the
assumption that the interaction has essentially compact support in time,
we can write
$$Z(A) = <0| (g(T),q(T))|0> . $$
The vacuum expectation value is given by a simple formula,  \cite{LM},
$$ <0| (g,q)|0>=\text{det}(aq^{-1}) $$ 
and therefore the parallel transport (with respect to the given local
trivialization) defines the phase of  the effective action $Z(A).$

Actually, the above discussion can be extended by a slight modification to the nonrenormalized time evolution
operators $U^I(t)$ in the interaction picture. This is important since there is a great freedom in the choice of the
family of unitary operators $T_A;$ basically $\tilde T_A = T_A Q_A$ is acceptable renormalization for any $Q_A \in U_{res}.$
We want to have a formula which does not depend on the choice of $T_A.$

\begin{lem} The time evolution in the interaction picture can be factorized as $$U^I(t) = (e^{itD_0} T_A^{-1} e^{-itD_0}) U^I_{ren}(t)$$
such that the off-diagonal blocks of $U^I_{ren}(t)$ are trace-class operators and $T_A -1$ is a pseudodifferential operator of order $-1.$
\end{lem}
 
{\bf Proof}
Denote by $H$ the Hilbert space of square integrable spinor fields on $\Bbb R^d,$ $D_0$ the 
free Dirac operator, and the grading $\epsilon = D_0/|D_0|.$

The time evolution of Dirac fermions in an external gauge potential $A$  is given as
$$ i\partial_t U(t) = (D_0 +V) U(t)$$
where $V=\alpha_k A_k$ is the 
interaction term.  According to \cite{LM}, we can choose an unitary operator 
$T_A$ which depends on $A$ and its time and space derivatives such that the renormalized
time evolution $U_{ren}(t) = T_A U(t)$ satisfies the differential equation
$$i\partial_t U_{ren}(t) = (D_0 +V_{ren}) U_{ren}(t),$$
where 
$$ V_{ren} = T_A^{-1} V T_A + T_A^{-1} i\partial_t T_A$$ 
is such that the commutator $[\epsilon, V_{ren}] \in L_2.$ Actually, it is possible to 
choose $T_A$ such that the commutator is even a trace-class operator, \cite{LM}.  The operator
$T_A$ is a pseudodifferential operator which differs from the unit operator by a pseudodifferential
operator of order $-1.$

Going to the interaction picture,
$ U^I_{ren} = e^{itD_0} U_{ren},$ we have 
$$ i\partial_t U^I_{ren} (t) = V_{ren}^I U^I_{ren}(t)$$
with $V^I_{ren} = e^{itD_0} V_{ren} e^{-itD_0}.$ Now the interaction $V^I_{ren}$ is an element of
the Lie algebra $\mathfrak{u}_{1}$ of the group $U_{1}$ and is a continuous function of
the time $t.$ It follows that the time evolution equation has a differentiable (in time) solution
$U^I_{ren}(t).$  

We can now factorize the original time evolution $U(t)$ as
$$U^I(t) = e^{itD_0} U(t) = e^{itD_0} T_A^{-1} e^{-itD_0} U^I_{ren}(t).$$ 
  
  $\square$

  We shall use a trace  extension on pseudodifferential operators, called the weighted trace by S. Paycha \cite{Pa}.  It gives the usual operator trace for trace-class
  operators, that is, on a compact manifold $M$ for pseudodifferential  operators of order strictly less than - dim$M.$ However, it is not cyclic for general pseudodifferential
  operators. First, one fixes a \it weight \rm as an elliptic invertible operator $Q$ of positive order $q.$ If $T$ is a pseudodifferential operator then the function
  $f(z) = \text{tr}\, Q^{-z} T$ is holomorphic in a half-plane $ Re(z) > \frac{1}{q} (\text{ord}(T) +  \text{dim}\, M)$ and can be continued to an analytic function in the neighborhood
  of $z=0$ with a simple pole at $z=0.$  The weighted trace of $T$ is defined as
  $$ \text{tr}_Q \, T = \lim_{z\to 0} \left( \text{tr} \,Q^{-z} T  - \frac{1}{qz} \text{Res}\, T \right)$$
  where $\text{Res}\, T$ is the Guillemin- Wodzicki residue of $T.$ Although the trace is not cyclic, the defect is given by the simple formula
  $$\text{tr}_Q [T,S] = - \frac{1}{q} \text{Res}\, T [\log Q, S].$$
  Although the logarithm of $Q$ is not a classical pseudodifferential operator, the commutator $[\log Q, S]$ is since its symbol is composed of the derivatives
  of the logarithm.   
  
  \bf Remark \rm One can define a $\Bbb C^{\times}$ bundle over the space of  all bounded invertible operators $g$ in $H=H_+\oplus H_-$  such that the block $a$ is 
  a Fredholm operator as in the case of the group $U_{res}:$ The total space is the set of pairs $(g,q)$ with $a-q$ trace-class, and the equivalence relation is
  $(g,q)\sim (g',q')$ for $g=g'$ and $\text{det}(qq'^{-1})=1.$ However, the total space is not a group.

  \begin{thm} Choosing the local section $q= a_g$ in the connection form $\phi = \text{tr}_Q  [ (dg g^{-1})_a - dq q^{-1}],$ for $g(t) = U^I(t)$ the operator under the trace
  is of the form $e^{itD_0} X e^{-itD_0} +Y$ where $X$ is a pseudodifferential operator of order $-1$ and $Y$ is a trace-class operator.  Thus choosing
  $Q= |D_0|$ the weighted trace $\text{tr}_Q$ is well-defined and equal to $\text{tr} Y + \text{tr}_Q X$ since $Q$ commutes with $e^{itD_0}.$  The result does not depend on the choice of splitting
  $(X,Y).$\end{thm} 
  
  {\bf Proof}  Write $g(t) = U^I(t) = g_1(t) g_2(t)$ with $g_1(t)= e^{itD_0} T_A^{-1} e^{-itD_0}$ and $g_2(t) = U^I_{ren} (t).$ Then 
  $ (dg(t) g(t) ^{-1})_a = V^I(t)_a$ and 
  $$da = P_+ dg P_+  = P_+ V^I(t) g(t)P_+  = (P_+ V^I P_+) a + (P_+ V^I P_-) (P_- g P_+) $$
  and therefore
  $$ da a^{-1} = P_+ V^I P_+  +(P_+ V^I P_-) (c a^{-1}).$$
  The first term on the right-hand-side cancels $(dg g^{-1})_a$ in $\phi$ whereas the second term is equal to 
  \begin{align*}  & (P_+ V^I P_-) (c_1 a_2 + d_1 c_2) (a_1 a_2 + b_1 c_2)^{-1}\\
 &= (P_+V^I P_-)(c_1 a_1^{-1} +d_1 c_2 a_2^{-1} a_1^{-1} )( 1+ b_1 c_2 a_2^{-1} a_1^{-1}) \equiv (P_+ V^I P_-)c_1 a_1^{-1} \text{ mod} L_1
  \end{align*}
  since $c_2 \in L_1.$  On the other hand, $c_1 = P_- e^{itD_0} T_A^{-1} e^{-itD_0} P_+$ and $a_1 = P_+(\dots) P_+.$ It follows 
  that the product $(P_+ V^I P_-) c a^{-1}$ is conjugate (by $e^{itD_0}$) to a pseudodifferential operator of order $-1,$
  modulo trace-class operators.
  
  Finally, let us assume that $e^{itD_0} X e^{-itD_0} + Y = e^{itD_0} X' e^{-itD_0} + Y'$ where also
  $Y'$ is trace-class and $X'$ is a pseudodifferential operator of order $-1.$ 
  Then $X-X'$ has to be also trace class  since $L_1$ is an ideal in the space of bounded
  operators. Since $\text{tr}_Q$ is a the standard trace for trace class operators we see that
  $\text{tr}_Q e^{itD_0}(X-X') e^{-itD_0} = \text{tr\,} e^{itD_0}(X-X') e^{-itD_0} = -\text{tr\,} (Y-Y')$ 
  which implies the uniqueness of the total trace. 
  
  $\square$
  
  \bf Remark \rm The above result can be formulated more generally: Let $\mathcal G$ be a space consisting of unitary operators $g= g_1 g_2$ with
  $g_1$ a conjugate of  a pseudodifferential operator by a unitary operator $R$ with $RQ=QR,$ and $g_2 \in U_{tr}.$ Let $t\mapsto g(t)$ be a differentiable path
  such that $dg g^{-1}= R(t) V(t) R(t)^{-1}$   and $g_1(t)= R(t) g_0(t) R(t)^{-1}$  for pseudodifferential operators $V(t), g_0(t).$ Then the formula (2), with tr replaced 
  by $\text{tr}_Q,$ defines a 
  connection in a $\Bbb C^{\times}$ bundle over $\mathcal G.$ However,  $\mathcal G$ is not a group.

 \section{Comparison with perturbation theory}
 
Together with formula (4) the above Theorem gives a method to compute the effective action $\log(Z).$
Let us consider the case of QED in four space-time dimensions.
Our method is nonperturbative,
but using the Dyson expansion
\begin{equation} g(t) = 1 -i\int_{-\infty}^t V_I(s) ds + (-i)^2\int_{t>s_1>s_2}  V_I(s_1)
V_I(s_2) ds_1 ds_2 + \dots  \end{equation}
for the time evolution operator in the interaction picture we get the
lowest ($A^2$ term) for $\log(Z),$
\begin{equation} \log(Z)= \int_{s>t} \text{tr} \,\, \pi_+ V_I(s) \pi_- V_I(t) \pi_+  dt \, ds.\end{equation}
In the case of QED, the terms of odd order are identically zero by parity invariance and the fourth order term is already
finite as a Feynman integral. So actually the second order term is the most interesting.

 So let us compute $\log(Z)$ to second  order in the interaction $A$ in the
case of QED, i.e., massive fermion coupled to a Maxwell potential.
Because of the unitarity relation $a^* a + c^* c=1$ the inverse $a^{-1}
\equiv a^*$ modulo terms of order $A^2$ since the the off diagonal blocks
of the time evolution $g(t)$ must contain the potential at least to
order one. For this reason the term $(\alpha - a^{-1}) a'$ in the phase does
not give contributions to the order $A^2.$ On the other hand, because of
unitarity, $\beta c'= c^* c' \equiv -bc'$ modulo terms of order higher than
two.

We shall use the integral representation

\begin{align} \frac{1}{2\pi i} & \int \text{tr}  \frac{\slashed{p} -m}{p^2 - m^2 +i\epsilon}
e^{ip_0 T}
dp_0 \nonumber \\ 
& =\gamma_0[\theta(T) \theta(h_0(\bold p)) e^{-iT\omega_p}
-\theta(-T) \theta(-h_0(\bold p)) e^{iT\omega_p}]
\end{align}
where $\omega_p=\sqrt{\bold p^2 +m^2},$  $\theta(x) =1$ for $x\geq 0$ and
$\theta(x)=0$ for $x<0$ and $h_0(\bold p)
= \gamma^0\gamma^k p_k +\gamma^0 m$ is the momentum representation for the free Dirac
hamiltonian. In the usual QED perturbation theory the second order effect is
given by the (diverging) Feynman integral
\begin{equation} \frac{1}{4\pi}\int\text{tr} \frac{\slashed{p} -m}{p^2 -m^2 +i\epsilon} \slashed{A}
(p-q)
\frac{\slashed{q}-m}{q^2 -m^2 +i\epsilon} \slashed{A}(q-p) d^4p\, d^4q      \end{equation}
with the Dirac notation $\slashed{X} = \sum\gamma_{\mu} X^{\mu}.$ 
By a Fourier transform of the potential in the time variable this integral
can be written as
$$\frac{1}{8\pi^2}\int \text{tr}  \frac{\slashed{p}-m}{p^2 -m^2 +i\epsilon} e^{is(p_0-q_0)}
\slashed{A}(s,\bold p
-\bold q) \frac{\slashed{q}-m}{q^2 -m^2+i\epsilon} e^{it(q_0-p_0)} \slashed{A}(t,\bold q
-\bold p) ds,dt\, d^4p\,d^4q.$$

Using the trick (7) above we can write the integral as

 \begin{eqnarray}  & -\frac12\int \text{tr} [\theta(s-t)\theta(h_0(\bold p)) e^{-i(s-t)
h_0(\bold p)} -\theta(t-s) \theta(-h_0(\bold p))
e^{-ih_0(\bold p) (s-t)}] \nonumber \\
& \times \slashed{A}(s,\bold p-\bold q) [\theta(t-s) \theta(h_0(
\bold q)) e^{-i(t-s)h_0(\bold q)}-\theta(s-t) \theta(-h_0(\bold q))e^{-i(t-s)
h_0(\bold q)}] \nonumber \\
& \times\slashed{A}(t,\bold q-\bold p)
e^{-ih_0(\bold q)(t-s)} ds\,dt\, d^3\bold p\,d^3\bold q.\end{eqnarray}

Since $\theta(T)\theta(-T)=0$ and $\theta(T)^2 =\theta(T)$ we get
$$ \int \text{tr} \,\theta(s-t) \pi_+ V_I(t) \pi_- V_I(s) \pi_+ ds\, dt $$
which is exactly the second order term in our geometric definition of
$\log(Z).$ Note that this discussion is formal in the sense that diverging
Feynman integrals are involved. However, we may apply some renormalization
method (for example, the family of $T_A$ operators described earlier,
to bring the time evolution to the group $U_{1}$) in order to make sense of
these integrals. But using our definition of parallel transport in terms of the
weighted trace $\text{tr}_{|D|}$ means that we are removing the logarithmic divergence
in the 1-loop diagram, which in the dimensional regularisation is the term $\frac{1}{z}\text{Res}.$
The limit $z\to 0 $ is just the limit $\epsilon \to 0$ in the dimension $4+\epsilon.$

\bibliographystyle{plain}
 
\end{document}